\documentclass[12pt]{article}
\usepackage{graphics}
\usepackage{graphicx}
\newcommand{\spc}{\hspace{0.22in}}
\newcommand{\spn}{\hspace*{0.22in}}

\newif\ifequationnumber
\def\eqswitchon{\equationnumbertrue
\@addtoreset{equation}{section}
\def\theequation{\arabic{section}.\arabic{equation}}
}
\def \ds {\displaystyle}

\def \ns {\normalsize}

\def \es {\enspace}
\def \ts {\thinspace}

\def \nms {\hspace*{-0.3cm}}

\title{
  \hfill{\ns hep-ph/0305062} \\ 
  \vspace*{-0.6cm}
  \hfill{\ns revised, June 2003} \\ 
  \vspace*{2.0cm}
  {\Large {\bf Destruction of Nuclear Bombs Using }} \\
  \vspace*{0.3cm}
  {\Large {\bf Ultra-High Energy Neutrino Beam}} \\
  \vspace*{0.5cm}
   {\Large --- dedicated to Professor Masatoshi Koshiba ---}
  \vspace*{1.0cm}
}
\author{
Hirotaka Sugawara\thanks{e-mail:sugawara@phys.hawaii.edu, \quad 
{\it Department of Physics and Astronomy,
University of Hawaii at Manoa, HI, USA}} \and
Hiroyuki Hagura\thanks{e-mail:hiroyuki.hagura@kek.jp, \quad
{\it Radiation Science Center of KEK, Tsukuba-shi, Japan}} 
\and Toshiya Sanami\thanks{e-mail:toshiya.sanami@kek.jp, \quad
{\it Radiation Science Center of KEK, Tsukuba-shi, Japan}}
}
\date{}
\begin{document}
\maketitle
%
%
\vspace{4cm} 
%
%
\begin{abstract}
We discuss the possibility of utilizing the ultra-high energy neutrino beam 
$(\simeq 1000 \ts {\rm TeV})$ to detect and destroy the nuclear bombs 
wherever they are and whoever possess them.
\end{abstract}
\newpage
%
\section{Introduction}
\spc Twentieth century physicists produced one of the most powerful weapons
on earth~\cite{oppenheimer} and they were used twice as an actual weapon with 
``Results Excellent.''\footnote{A coded message sent to President Harry 
S.~Truman from Richard Nelson, the youngest crewman of Enola Gay, who died 
February~7, 2003 of complication of emphysema (New York Times, February~7, 
2003).} The number of countries which possess or will
possess nuclear weapons could increase in spite of the existence of Treaty
on the Non-Proliferation of Nuclear Weapons (NPT). There is no guarantee that 
these countries which already possess nuclear weapons always behave 
humanistically. Arms control negotiations may stabilize the world temporarily 
but, again, there is no guarantee that the long lasting peace on earth will 
come true in the future. We discuss in this article a rather futuristic but 
not necessarily impossible technology which will expose the possessors of 
nuclear weapons in an extreme danger in some cases. 

  Our basic idea is to use an extremely high energy neutrino beam which 
penetrates the earth and interacts just a few meters away from a potentially 
concealed nuclear weapon. The appropriate energy turns out to be about 
1000~TeV. This is the energy where the neutrino mean free path becomes 
approximately equal to the diameter of the earth. The neutrino beam produces a 
hadron shower and the shower hits the plutonium or the uranium in the bomb and 
causes fission reactions. These reactions will heat up the bomb and either melt
it down or ignite the nuclear reactions if the explosives already surround the 
plutonium. We will calculate the intensity of the neutrino beam required and 
the duration of time which the whole process will take place for a given 
intensity.

  We emphasize that the whole technology is futuristic and the reason should
be clear to all the accelerator experts. Actually, even the simplest prototype
of our proposal, i.e.~the neutrino factory of GeV range needs substantial 
R~\&~D work. We also note that a 1000~TeV machine requires the accelerator
circumference of the order of 1000~km with the magnets of 
$\simeq 10~{\rm Tesla}$ which is totally ridiculous. Only if we can invent a
magnet which can reach almost one order of magnitude higher field than the 
currently available magnet, the proposal can approach the reality. Even if
it becomes the reality, the cost of the construction is of the order of or more
than 100~billion~US\$. Also we note that the power required for the operation 
of the
machine may exceed 50~GW taking the efficiency into account. This is above the 
total power of Great Britain. This implies that no single country will be able 
to afford the construction of this machine and also the operation time must be
strictly restricted. We believe the only way this machine may be built is when
all the countries on earth agree to do it by creating an organization which may
be called the ``World Government'' for which this device becomes the means of
enforcement.

  Section~\ref{sect:roughEstmtn} gives a rough estimation and 
section~\ref{sect:simulation} deals with a computer simulation using a 
Monte-Carlo generator MCNPX~\cite{MCNPX}. 
Section~\ref{sect:conclusion} gives a conclusion with various comments.
In addition, we give the calculation of the mean free path of neutrino inside 
the earth in appendix~\ref{appd:meanFP}. In appendix~\ref{sect:acclrtrNewTch},
we also describe a possible accelerator scheme. 
\vspace*{0.3cm}
%
\section{Rough estimation}\label{sect:roughEstmtn}
\begin{figure}[h]
\vspace{0.4cm}
\begin{center}
\setlength{\unitlength}{1cm}
\begin{picture}(13,5.5)(0,0)
\put(0.5,0.3){\includegraphics[width=12cm]{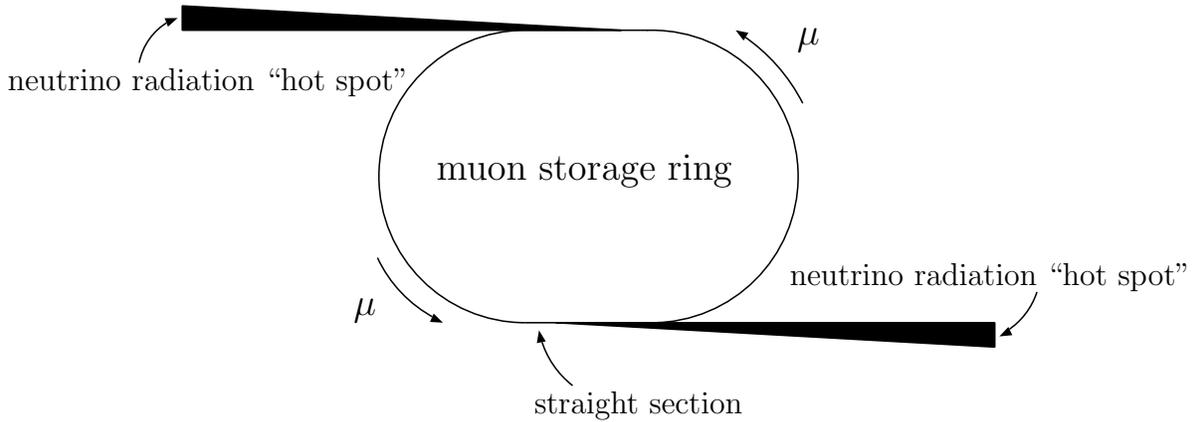}}
\put(4.5,3.1){\large muon storage ring} 
\put(3.4,1.3){\large $\mu$} \put(9.3,4.9){\large $\mu$}
\put(5.8,0.0){straight section}
\put(9.2,1.7){neutrino radiation ``hot spot''}
\put(-1.2,4.3){neutrino radiation ``hot spot''}
\end{picture}
\end{center}
\vspace*{-0.8cm}
\caption
{
{\sl Neutrino radiation from a race track shaped muon storage ring.
The decay of muons will produce the neutrino radiation emanating out 
tangentially everywhere from the ring. In particular, the straight sections in 
the ring will cause radiation ``hot spots'' where all of the decays line up 
into a pencil beam.}
}
\label{fig:muonRing}
\vspace*{0.4cm}
\end{figure}

\spc The neutrino beam is already hazardous at the energy of several TeV as 
have been analyzed by B.~J.~King~\cite{King} and N.~V.~Mokhov and 
A.~Van~Ginneken~\cite{MokhovGinneken} in connection with the study of the muon
colliders. If one constructs a race track shaped muon storage ring as shown in
fig.~\ref{fig:muonRing}, most of the muons decay into the two opposite
directions of the straight sections. These two directions are the most 
hazardous directions (``hot spots'') but the circular parts also emit the 
neutrinos which may also be hazardous in the vicinity of the storage ring.
%
%

\begin{figure}
\begin{center}
\setlength{\unitlength}{1cm}
\begin{picture}(14.0,5.0)(0,0)
\put(1.0,0.0){\includegraphics[width=12cm]{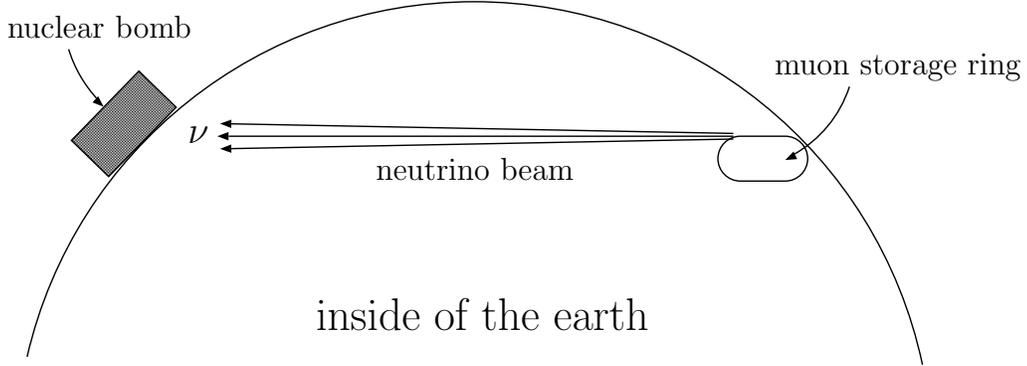}}
\put(0.8,4.4){nuclear bomb} \put(11.0,3.9){muon storage ring}
\put(3.2,3.0){\large $\nu$}
\put(5.7,2.5){neutrino beam}
\put(4.9,0.5){\Large inside of the earth}
\end{picture}
\end{center}
\vspace*{-0.8cm}
\caption
{
{\sl Neutrino beam is aimed at the nuclear bomb that is placed on the 
opposite side of the earth. The beam is emitted downstream from one of the 
straight sections of the muon storage ring (see fig.~\ref{fig:muonRing}), and 
reaches the bomb after passing through the inside of the earth.}
}
\label{fig:lineThrghEarth}
\vspace{0.3cm}
\end{figure}

\begin{figure}[h]
\vspace{0.0cm}
\begin{center}
\setlength{\unitlength}{1cm}
\begin{picture}(11.9,9.0)(0,0)
\put(0.5,0.0){\includegraphics[width=10.9cm]{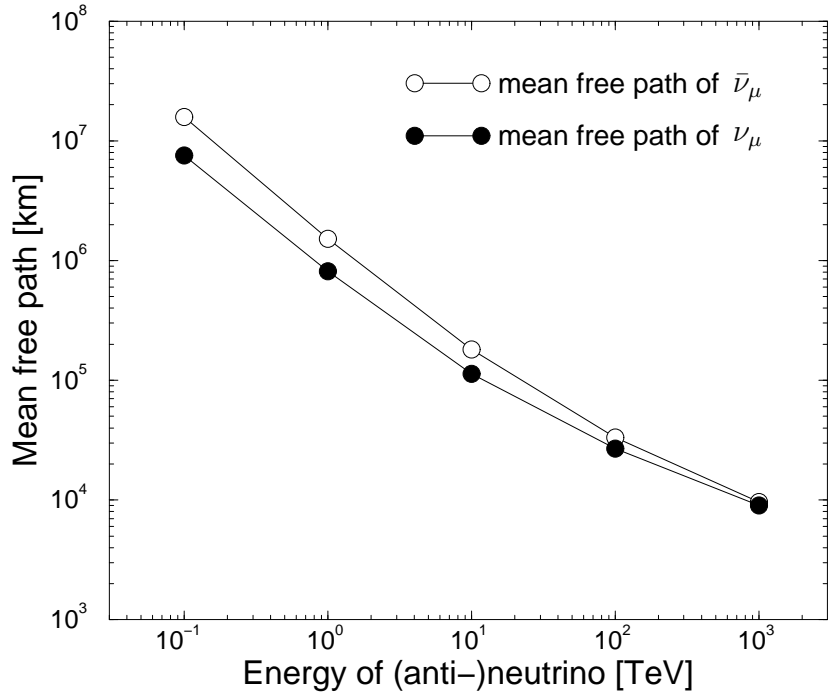}}
\put(10.15,8.2){$\bar{\nu}_{\mu}$} \put(10.15,7.55){$\nu_{\mu}$} 
\end{picture}
\end{center}
\vspace*{-1.2cm}
\caption
{
{\sl Mean free path of (anti-)neutrino vs.~its energy. This is calculated 
under the assumption that in this energy region the deep inelastic cross 
sections dominate. For the detail of the calculation, see 
Appendix~\ref{appd:meanFP}.}
}
\label{fig:mfpVsEng}
\vspace{0.5cm}
\end{figure}

  Here we would like to consider a situation where one of the straight lines
is directed toward the nuclear bomb which is located somewhere on the 
opposite side of the earth (fig.~\ref{fig:lineThrghEarth}).
We must choose the energy of the neutrino beam in such a way that the mean
free path of the neutrino is compatible to the diameter of the earth.
Fig.~\ref{fig:mfpVsEng} shows the mean free path of (anti-)neutrino vs.~its 
energy calculated assuming that the deep inelastic cross section dominates in 
the relevant energy region\footnote{See appendix~\ref{appd:meanFP} for the 
detailed calculations of the mean free paths shown in 
fig.~\ref{fig:mfpVsEng}.}.
From fig.~\ref{fig:mfpVsEng} we conclude that the energy of the neutrino beam
must be about 1000~TeV to have approximately single interaction before the
neutrino beam hits the bomb. The size of the beam at 
the point of the bomb is given by
\begin{eqnarray*}
r = \frac{\ds m_{\mu} c^2}{\ds E_{\nu}} d \ts \simeq 
\frac{\ds 0.1~{\rm (GeV)} \times 10^7 ~{\rm (m)}}{\ds 10^6~{\rm (GeV)}} 
= 1~{\rm (m)} \es,
\end{eqnarray*}
where $m_{\mu}$ and $c$ stand for the muon mass and the speed of light, and
$d$ is the distance from the muon storage ring to the position of the bomb 
which we take to be the diameter of the earth $(\simeq 10^7 \ts {\rm m})$. The 
beam spread due to the transverse momentum of the beam is negligible at this 
energy if the current value of the ionization cooling of $P_{t} = 
1~{\rm (MeV)}$ is adopted. The range of the neutrino is $10^7$ meters and the 
effective neutrino interaction is restricted within a few meters away from the 
bomb because of the interaction range of the hadrons. Therefore, the 
probability of getting an effective reaction from the beam is $1/10^{7}$. 
As a result, the energy deposit from the beam for the unit area~$({\rm m}^2)$ 
is given by
\begin{eqnarray}
E_{\rm dep} = 10^{15} \times 10^{-7} \times I = 10^{8} I \es 
{\rm (eV/sec \cdot m^2)} \es,
\label{eq:Edep}
\end{eqnarray}
where $I$ stands for the neutrino beam intensity. For example, we get 
\begin{eqnarray}
E_{\rm dep} \simeq 1000 \es {\rm (Joule/sec \cdot m^2)} \quad
\mbox{for the intensity of } I = 10^{14}~{\rm (1/sec)} \ts.
\label{eq:depToIntnsty}
\end{eqnarray}
This is equivalent to about $1~{\rm S}_{\rm V}/{\rm sec}$. We note that this 
value of the radiation dose is very large, compared with the U.S.~Federal 
off-site limit of $1~{\rm mS}_{\rm V}/{\rm year}$.\footnote{The unit 
${\rm S}_{\rm V}$ corresponds, in alternative units, to $100~{\rm rem}$.}
%
%

  The above estimation can be summarized in the following formula:
\begin{eqnarray}
E_{\rm dep} = E_{\nu} \cdot \left(\frac{\ds R_h}{\ds R_{\nu}}\right) 
\cdot I \cdot \left(\frac{\ds m_{\mu} c^2}{\ds E_{\nu}} d \right)^{-2}
~{\rm (eV/sec\cdot m^2)} \es,
\label{eq:Edepformula}
\end{eqnarray}
where $R_h$ and $R_{\nu}$ stand for the average hadron and the neutrino mean
free paths. $R_{\nu}$ is proportional to $E_{\nu}^{-1}$ below 1000~TeV 
corresponding to $\sigma^{\rm tot}_{\nu}~({\rm cm}^2) \sim 10^{-38} E_{\nu}~
({\rm GeV})$ which leads to 
\begin{eqnarray*}
E_{\rm dep} \sim E_{\nu}^4 \es.
\label{eq:EdepEstmt}
\end{eqnarray*}
$E_{\rm dep}$ drops sharply to 0.1~Joule/s for 100~TeV neutrino energy. It is,
therefore, rather crucial to keep the energy as high as 1000~TeV.
%
%
\begin{figure}
\begin{center}
\setlength{\unitlength}{1cm}
\begin{picture}(11,8.5)(0,0)
\put(1.1,0.0){\includegraphics[width=9cm]{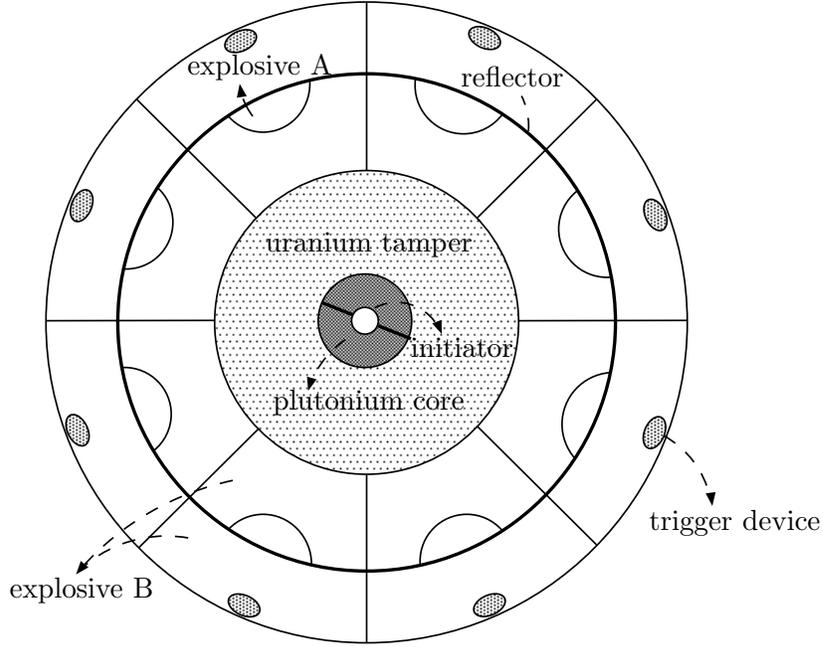}}
\put(4.1,5.3){\small uranium tamper} \put(6.03,3.9){\small initiator}
\put(6.7,7.5){\small reflector} \put(3.06,7.66){\small explosive A}
\put(4.2,3.2){\small plutonium core} \put(9.2,1.6){\small trigger device} 
\put(0.7,0.7){\small explosive B}
\end{picture}
\end{center}
\vspace{-0.8cm}
\caption
{
{\sl A model for the plutonium bomb of implosion type~{\rm \cite{RCJHHMW}}. 
The whole profile of the bomb is in the shape of a spherical body.}
}
\label{fig:mdlPltnmBmb}
\vspace{0.5cm}
\end{figure}

  To proceed further we need to know a little about the structure of the 
nuclear bomb. Since no official information is available to us, we rely on 
popular books~\cite{Server,RCJHHMW} and unclassified 
papers~\cite{unclsfdPapers} on the subject. 
As a possible model for the bomb we consider a 10~kg ball of 
${}^{239}{\rm Pu}$ which has the critical mass of 15~kg, surrounded by the 
${}^{238}{\rm U}$ tamper, the reflector and the explosive material 
(fig.~\ref{fig:mdlPltnmBmb}). We also consider a system without explosive 
material surrounding the plutonium ball since we have no way to know how these 
bombs are stored. A crucial parameter in the former case is the number of 
fissions in the system which provides the temperature rise enough to ignite 
the surrounding explosives. If we assume that the explosive has the ignition 
temperature of $300~^{\circ}{\rm C}$ (TNT~(TriNitroToluene) has its ignition 
temperature $210~^{\circ}{\rm C}$), the number of fissions, $N_{\rm fission}$, 
required turns out to be about $10^{16}$ per $10~{\rm kg}$ of
plutonium~\cite{unclsfdPapers}. One order of magnitude larger 
value $10^{17}$ should be enough to melt down the system in the latter case. 

  Let us estimate how much time it takes for this process to happen when the
energy deposit from the neutrino beam is given by eq.~(\ref{eq:Edep}).
Since the tamper ${}^{238}{\rm U}$ can also be regarded as the source of the
fission when the neutron energy is as high as 10~MeV, which is the typical
energy, $\epsilon_{\rm sp}$, of the spallation neutrons, we take the 
area exposed to the hadron shower to be $0.1~({\rm m}^2)$. In this case 
the total energy deposit in the bomb fission system is
\begin{eqnarray*}
E_{\rm dep}^{\rm T} = 0.1 \times 10^8 I = 10^{21}~{\rm (eV/sec)} 
\quad {\rm for} \es I = 10^{14}~{\rm (1/sec)}\es.
\end{eqnarray*}
The number of spallation neutrons, therefore, is equal to
\begin{eqnarray*}
n_{\rm sp} = \frac{\ds E_{\rm dep}^{\rm T}}{\ds \epsilon_{\rm sp}}
= \frac{\ds 10^{21}}{\ds 10^{7}} = 10^{14}~{\rm (1/sec)} \es.
\end{eqnarray*}
If we assume a spallation neutron causes a single fission and none are lost,
the time required to ignite the explosive is 
\begin{eqnarray*}
\frac{\ds N_{\rm fission}}{n_{\rm sp}} = \frac{\ds 10^{16}}{\ds 10^{14}}
= 100~({\rm sec}) \es,
\end{eqnarray*}
and the melt down time is about 1000~sec. The next question is whether 
we are going
to have a full fledged explosion or a kind of ``fizzle explosion'' in the case
when the explosive sets ignited. The problem was studied in the appendix of
ref.~\cite{unclsfdPapers} by F.~von Hippel and E.~Lyman. This analysis shows 
that the neutron from the spontaneous fission which contaminates the 
${}^{239}{\rm Pu}$ system gives the probability of the occurrence of ``fizzle 
explosion'' given by
\begin{eqnarray}
P = 1 - e^{-c n_{\rm spon}} \es,
\label{eq:probFzzlEx}
\end{eqnarray}
where $n_{\rm spon}$ is the number of neutrons from the spontaneous emission 
and $c$ is a certain constant. We can replace this number $n_{\rm spon}$
by the number of spallation neutrons caused by the neutrino beam, which is 9 
order of magnitude larger than the number of neutrons from the spontaneous 
emission of  ${}^{240}{\rm Pu}$. Therefore, the probability of the ``fizzle 
explosion'' is practically equal to 1 in this case. This results in an energy 
yield of the explosion by the neutrino beam to be about 3\% of the 
full explosion.
\vspace*{0.1cm}
%
\section{Simulation}\label{sect:simulation}
\begin{figure}[h]
\vspace{0.3cm}
\begin{center}
\setlength{\unitlength}{1cm}
\begin{picture}(11,7.0)(0,0)
\put(1,0){\includegraphics[width=9cm]{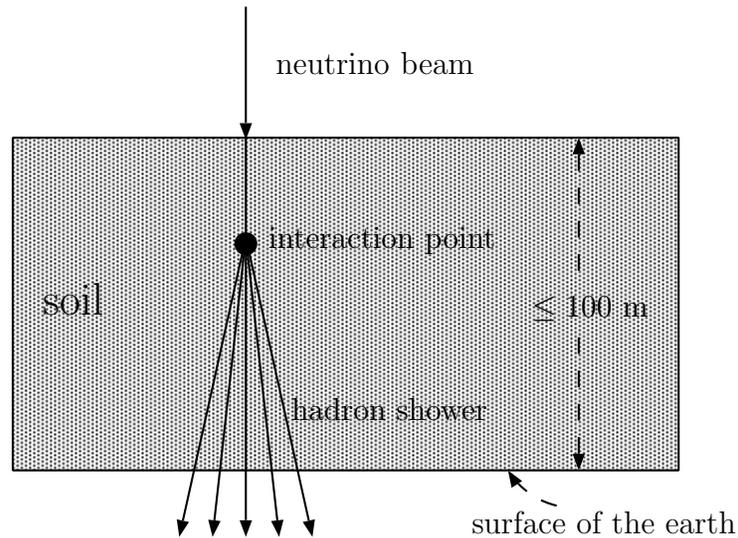}}
\put(4.6,6.2){neutrino beam}
\put(4.5,3.9){interaction point} \put(8.0,2.98){$\le 100$~m}
\put(1.5,3.0){\Large soil} \put(4.8,1.6){hadron shower}
\put(7.2,0.1){surface of the earth}
\end{picture}
\end{center}
\vspace*{-0.8cm}
\caption
{
{\sl Hadron shower arising near the target bomb.
The neutrino beam passing through the soil interacts with
nuclei near the surface of the earth, resulting in a hadron
shower in a place a few meters close to the bomb.}
}
\label{fig:systmErth}
\vspace*{0.0cm}
\end{figure}

\spc Having discussed the rough estimation, let us now turn to numerical
simulations to study the system in a more precise way.
We divide our simulation procedure into two parts: The first part deals
with a neutrino beam and its 
development into a hadron shower (see fig.~\ref{fig:systmErth}). The second 
one follows the first one to calculate the nuclear reactions in the target bomb
(see fig.~\ref{fig:pltnmSys}). We give detailed discussions of the 
two parts separately in the subsequent sections~\ref{sect:incdntNtrn} and
\ref{sect:rctnsCore}.
%
%
\subsection{Incident neutrino beam and hadron shower}\label{sect:incdntNtrn}
\spc The first part is to start from a given neutrino beam of certain energy 
and intensity. We simulate the process of the neutrino beam hitting a target 
nucleus in the soil and follow the development of a hadron shower initiated by 
the neutrino interaction, as shown in fig.~\ref{fig:systmErth}. 
The former process can be 
simulated by a generator HERWIG~\cite{HERWIG}, in which we can include 
processes with the incident neutrino beam, such as $\nu + p 
\mbox{ (or $n$)} \to {\rm hadrons} + {\rm leptons}$. In the latter process, 
subsequently, we simulate the interactions of the hadron shower
with nuclei of the soil by using another generatorS GEANT4~\cite{GEANT} and
MARS~\cite{MARS}. The purpose of this part is to 
obtain the multiplicity of the shower when the shower is going out of the 
earth. The neutrino interaction which occurs near the surface of the earth is 
relevant. We consider, therefore, a system which is shown in 
fig.~\ref{fig:systmErth}. 

The result of this part will appear in a separate publication~\cite{FTPR}.
%
\subsection{Nuclear reactions inside the target}\label{sect:rctnsCore}
\begin{figure}[h]
\begin{center}
\setlength{\unitlength}{1cm}
\begin{picture}(12,6.4)(0,0)
\put(1,0.0){\includegraphics[width=10cm]{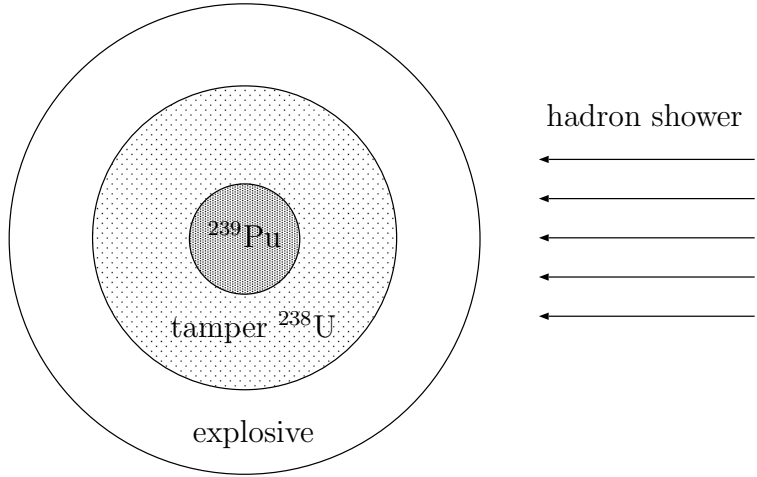}}
\put(3.7,3.1){${}^{239}{\rm Pu}$} \put(8.2,4.7){hadron shower}
\put(3.2,1.9){tamper ${}^{238}{\rm U}$}
\put(3.5,0.5){explosive}
\end{picture}
\end{center}
\vspace{-0.8cm}
\caption
{
{\sl A hadron shower going into the plutonium bomb. It will
induce the fission reactions inside the plutonium system and cause
the temperature increase as a result.}
}
\label{fig:pltnmSys}
\vspace{0.4cm}
\end{figure}

\spc The second part of our simulation is to calculate the temperature
increase of the plutonium system caused by the hadron shower. We consider a 
system shown in fig.~\ref{fig:pltnmSys}. Our calculation of this part is 
carried out using the MCNPX code\footnote{MCNPX has been used 
extensively in nuclear reactor physics and its applications, and developed 
for a long time since 1940's~\cite{MCNPX}. Furthermore, many physicists and 
programmers are still developing it for improvement even at present.}.

\begin{figure}
\begin{center}
\setlength{\unitlength}{1cm}
\begin{picture}(12,8.0)(0,0)
\put(1,0.0){\includegraphics[width=9.6cm]{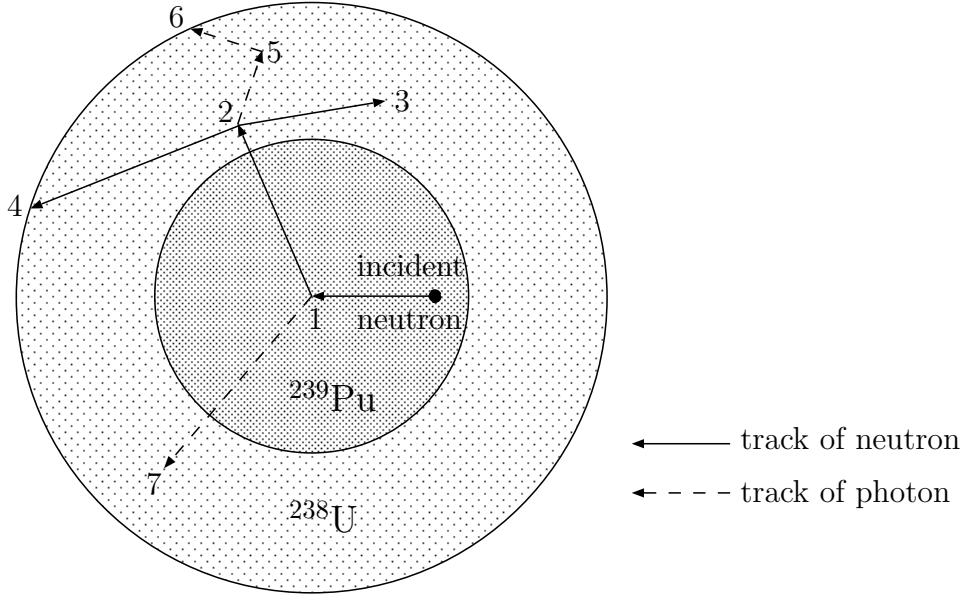}}
\put(5.6,4.3){incident} \put(5.6,3.6){neutron} \put(4.95,3.6){1}
\put(6.1,6.5){3} \put(3.75,6.36){2} \put(4.4,7.15){5} 
\put(2.8,1.4){7} \put(0.95,5.1){4} \put(3.1,7.62){6}
\put(4.7,2.5){\large ${}^{239}{\rm Pu}$} 
\put(4.7,0.90){\large ${}^{238}{\rm U}$}
\put(10.7,2.0){track of neutron} \put(10.7,1.3){track of photon}
\end{picture}
\end{center}
\vspace{-0.8cm}
\caption
{
{\sl A history of a neutron incident on the ${}^{239}{\rm Pu}$ core that 
can undergo nuclear fission. 1.~Neutron scattering and photon production in
the core. 2.~Fission and photon production in the ${}^{238}{\rm U}$ tamper. 
3.~Neutron capture in the tamper. 4.~Neutron leakage out of the tamper.
5.~Photon scattering in the tamper. 6.~Photon leakage out of the tamper. 
7.~Photon capture in the tamper.}
}
\label{fig:mcnpxSmltnPrcss}
\vspace{0.4cm}
\end{figure}

  In order to simplify the system and save the computation time, we replace
the parallel hadron shower of fig.~\ref{fig:pltnmSys} by a neutron source 
which is situated at the center of the ${}^{239}{\rm Pu}$ core, as shown in 
fig.~\ref{fig:mcnpxSmltnPrcss}. We also assume that the incident neutron of 
its energy $1~{\rm GeV}$ starts from a point inside the core. 

  To make clear the algorithm of the MCNPX code, we illustrate
a simple example of nuclear reactions in fig.~\ref{fig:mcnpxSmltnPrcss}, where 
the first collision occurs at event 1 in the Pu core. 
The neutron is scattered in the direction shown,
which is selected randomly from the physical scattering distribution. A photon
is also produced and is temporarily stored, or banked, for later analysis. At 
event 2, fission occurs, resulting in the termination of the incoming neutron
and the birth of two outgoing neutrons and one photon. One neutron and the 
photon are banked for later analysis. The first fission is captured at event 3
and terminated. The banked neutron is now retrieved and, by random sampling,
leaks out of the core at event 4. The fission-produced photon has a collision
at event 5 and leaks out at event 6. The remaining photon generated at event 1
is now followed with a capture at event 7. Note that MCNPX retrieved banked 
particles such that the last particle stored in the bank is the first particle
taken out. 

  This neutron history in the core and the tamper
is now complete. As more and more such 
histories are followed, the neutron and photon distributions become better
known. The quantities of interest, such as the total energy arising in the
reactions, are tallied along with estimates of the statistical precision of 
the results. Hence, after repeating the similar calculations, we could obtain 
the average value, $\epsilon_{\rm fission}$, of the fission energy deposition:
\begin{eqnarray*}
\epsilon_{\rm fission} = 0.6260 \pm 0.0032~{\rm (MeV/g)}\es.
\end{eqnarray*}
This is the contribution on the average from one incident neutron. Thus, if 
we prepare $N_{\rm in}$ neutrons incident on the ${}^{239}{\rm Pu}$ core,
the increase in temperature can be estimated as
\begin{eqnarray}
\Delta T = \frac{\ds N_{\rm in} \ts
\epsilon_{\rm fission}}{\ds C_{\rm Pu}} = 0.9547 \times 10^{-12} \ts
N_{\rm in}~{\rm (K)} \es,
\label{eq:incTmprtr}
\end{eqnarray}
where $C_{\rm Pu}$ is the specific heat of ${}^{239}{\rm Pu}$, whose numerical
value is given by
\begin{eqnarray*}
C_{\rm Pu} = \frac{\ds 4.186~{\rm (J/cal)} \times (6.0/239)~{\rm 
(cal/g\cdot K)}}{\ds 1.602 \times 10^{-13}~{\rm (J/MeV)}}
= 6.557 \times 10^{11}~{\rm (MeV/g\cdot K)} \es.
\end{eqnarray*}
Therefore, in order to obtain a temperature increase $\Delta T = 
250~{\rm (K)}$, which corresponds to the ignition temperature of TNT,
the total number of the incident neutrons should be
\begin{eqnarray}
N_{\rm in} = \frac{\ds 250 }{\ds 0.9547 \times 10^{-12}} 
= 2.619 \times 10^{14} \es.
\label{eq:incdntNmbr}
\end{eqnarray}
This value of $N_{\rm in} \sim O(10^{14})$ is close to the estimated value
in section~\ref{sect:roughEstmtn}, and it is not unrealistic in the future
technology of the muon colliders.

  To verify our results of simulation, the experiment can be devised in
which a hadron beam of GeV energy hits a ball of plutonium (not necessarily
of weapon grade) and increases its temperature. The experiment is 
similar to the one which is performed to study the target material for the
neutron spallation sources.
\vspace*{0.1cm}
%
\section{Conclusion}\label{sect:conclusion}
\spc We have shown that it is possible to eliminate the nuclear bombs from the
surface of the earth utilizing the extremely high energy neutrino beam. When
the neutrino beam hits a bomb, it will cause the fizzle explosion with 3\% of
the full strength. It seems that it is not possible to decrease the magnitude
of the explosion smaller than this number at this stage.
It is important to decrease this number to destroy bombs safely.
We are not sure what this means when the plutonium or uranium is used to ignite
the hydrogen bomb. We may just break the bomb or may lead to a full explosion.
The whole process takes a matter of a few minutes in the case considered in
this paper although, of course, it depends on the intensity of the neutrino
beam. When the bombs are stored in the form of plutonium ball separated from 
the explosives, what we can do is to melt them down or vapor them away. 
It takes substantially longer time for this process to occur. 

%
%

 To justify the above statements we performed a detailed simulation calculation
and the part of its results is explained in this paper although the full 
content will be published later.
   After the high energy neutrino beam passes through the soil, it causes 
a hadron shower near the surface of the earth, and subsequently, neutrons in 
the shower will strike the ${}^{239}{\rm Pu}$ core. In order to estimate the 
number of the incident neutrons which is large enough to make the temperature 
of the TNT surrounding the core increase to its ignition one, we have carried 
out the numerical simulations using MCNPX under the simplified conditions. 
As a consequence, we obtained the value of $N_{\rm in} \sim O(10^{14})$.
This value is consistent with the estimation obtained roughly, 
and it is expected to be realistic in the future technology. 

  We utilize 1000~TeV\footnote{Actually, the neutrino mean free path which 
leads us to consider 1000~TeV is not quite accurate. As is mentioned in 
Appendix~\ref{appd:meanFP}, we did not include the contribution of the heavier
quarks. In addition, we did not take into consideration the ``neutrino 
transport theory''~\cite{Naumov} at all here. 
The inclusion of there effects on the deep inelastic cross-sections will
lead to the mean free path which is almost 1/3 of the value we have used in
this paper. This change will lead to the energy of 300~TeV and, therefore, we
need 27 times higher intensity than considered in 
section~\ref{sect:roughEstmtn}. Of course, targeting the bomb becomes
much easier.}
 neutrino beam for our purposes and we do not have the 
technology yet to produce such a high energy neutrino. We may start an R~\&~D 
now and proceed step by step. Yet it may take even an order of a century to 
achieve the goal.
\vspace*{0.5cm}

  We describe below a possible scenario for the whole project:
\begin{itemize}
\setlength{\itemsep}{-2pt}
\item[(1)] First, we should construct a neutrino factory which could have 
substantially lower energy than even 1~TeV. The purpose is to fully understand
the properties of the neutrino including mass, mixing angles, CP violating 
phase, Majorana property and the interactions with other particles~\cite{KP}.
\item[(2)] The next step is to construct a muon collider in the multi-TeV 
energy range.
The energy should be beyond 10~TeV. These two steps
still require a fair amount of R~\&~D but we believe that they are on the 
straightforward extension of the currently available technology. 
\item[(3)] The third step is to construct a muon collider
 of more than 100~TeV energy and to reach even 1000~TeV. We do not have
the technology yet for this kind of machine. We may need a magnet with one to
two orders of magnitude higher field than the currently available one to 
construct a machine of a reasonable scale. A completely new approach may be
necessary to make this possible.
\item[(4)] If the third step becomes real, then the fourth step is to actually
build the 1000~TeV muon collider with the movable straight sections.
\end{itemize}

  We want to emphasize the importance of realizing the first two steps since 
the technology is within the reach and its contribution to the basic science
is enormous. The study may even show that the neutrino interaction increases
more rapidly with the energy owing to the extra-dimension as in some 
models~\cite{KTZ}.
In that case we need not go as far as 1000~TeV to achieve our goal. The last
two steps require tremendous amount of effort in developing the necessary 
technology. It is also true that fair amount of financial and human resources
will have to be introduced to accomplish the last two steps.

  The neutrino beam could also be used to detect the nuclear bombs with much 
less energy and with much less intensity. The necessary technology is the 
detection of the fission products from a reasonable distance. It could be 
rather difficult if the bombs are stored in a deep underground location.

  Another useful application of high energy neutrino beam is to the study of
the inner structure of the earth~\cite{RGWC}. 
We may not need the neutrino energy to be as
high as 1000~TeV in this case. The detailed study is being performed on this
subject and we will describe it in a forthcoming publication.

  We are certainly aware of the fact that this kind of device can not only
target the nuclear bombs but other kinds of weapons of mass destruction and 
also, unfortunately, any kind of living object including human. But we should
emphasize that the device itself is not a weapon of mass destruction. 
The reason is as follows: The calculation in section~\ref{sect:roughEstmtn}
and section~\ref{sect:simulation} shows that it takes 1~second for this device
to cover a $1~{\rm m}^2$ with the radiation dose of $1~\mbox{S}_{\rm V}$.
It takes more than a year to cover the area of $10~{\rm km}^2$ with this value
of dose per unit area. It is extremely unlikely that no measure is taken after
a few minutes of exposure of this kind. Moreover, as is emphasized in the
introduction and also in the appendix~\ref{sect:acclrtrNewTch}, the 
construction cost and the power required for the operation make it almost 
impossible for even the richest country to build and operate it all by itself.
We strongly object to the ungrounded worry that this kind of device, even its
downgraded version could be used by certain irresponsible organization as a
weapon of mass destruction. On the contrary, we sincerely hope that our
proposal will motivate and stimulate the revival of the old idea of 
``World Government'' which has so far been discarded as unrealistic.

  Lastly we would like to point out that at least the first two steps described
in this section have nothing to do with the weapon research. They belong to the
most fundamental scientific research activities. The suitable organizational
structure to perform such a research, therefore, is through the world-wide
collaboration.
  Another worry could be expressed on the neutrino hazard around the machine.
It depends crucially where one builds the machine. The concrete proposal 
explained in appendix~\ref{sect:acclrtrNewTch} has two hazardous planes and two
dangerous ($P_3 P_4$ and $Q_3 Q_4$) directions. "No fly zones" 
should be set to avoid 
these hazardous regions. The duration time of an operation should be minimized 
for the security reason and also for the reason of power consumption.
\clearpage
%
%
\noindent{\Large {\bf Appendices}}
\appendix
\section{Mean free path of (anti-)neutrino}\label{appd:meanFP}
\spc This section is devoted to the derivation of the mean free path of 
(anti-)neutrino passing through the inside of the earth. T.~Abe, who is on the 
staff of Accelerator Laboratory of KEK, has mainly contributed to the following
calculations and actually made fig.~\ref{fig:mfpVsEng} under some assumptions. 
Here we discuss the assumptions in some detail and further derive the mean 
free paths in a numerical way.

  First, we assume that the deep inelastic scattering, including the
contributions of both the charged and neutral currents, dominates in the
energy region multi~TeV $\sim$ $10^3$~TeV; the relevant cross sections are
given at the tree level~\cite{AichisonHey} as
\begin{eqnarray}
\frac{\ds d^2 \sigma_{\rm cc}^{\nu}}{\ds dx dy} 
\nms&=&\nms 
\frac{\ds G_{\rm F}^2 m_{\rm N} E_{\nu} x}{\ds \pi} 
\left( \frac{\ds M_W^2}{\ds -q^2 + M_W^2} \right)^2
\Biggl\{ Zd(x) + Nu(x) 
+ (1-y)^2 \left(Z\bar{u}(x) + N\bar{d}(x)\right) \Biggr\} \ts, 
\label{eq:DIchrgd1} \\
\frac{\ds d^2 \sigma_{\rm cc}^{\bar{\nu}}}{\ds dx dy}
\nms&=&\nms 
\frac{\ds G_{\rm F}^2 m_{\rm N} E_{\nu} x}{\ds \pi} 
\left( \frac{\ds M_W^2}{\ds -q^2 + M_W^2} \right)^2
\Biggl\{ Z\bar{d}(x) + 
N\bar{u}(x) + (1-y)^2 \Bigl(Z u(x) + Nd(x)\Bigr) \Biggr\} \ts,
\label{eq:DIchrgd2} \\
\frac{\ds d^2 \sigma_{\rm nc}^{\nu}}{\ds dx dy}
\nms&=&\nms 
\frac{\ds G_{\rm F}^2 m_{\rm N} E_{\nu} x}{\ds \pi} 
\left( \frac{\ds M_Z^2}{\ds -q^2 + M_Z^2} \right)^2 \nonumber \\
\nms&\times&\nms
\Biggl[ (1-y)^2 \Biggl\{ 
\left(\frac{\ds 2}{\ds 3}\sin^2\theta_W\right)^2 \Bigl(Zu(x) + 
Nd(x)\Bigr) + \left(\frac{\ds 1}{\ds 3}\sin^2\theta_W\right)^2
\Bigl(Zd(x) + Nu(x)\Bigr) \nonumber \\
\nms&+&\nms \left(- \frac{\ds 1}{\ds 2} + \frac{\ds 2}{\ds 3}\sin^2
\theta_W\right)^2 \left(Z\bar{u}(x) + N\bar{d}(x)\right) + 
\left(\frac{\ds 1}{\ds 2} - \frac{\ds 1}{\ds 3}\sin^2\theta_W\right)^2
\Bigl(Z\bar{d}(x) + N\bar{u}(x)\Bigr) \Biggr\} \nonumber \\
\nms&+&\nms  \Biggl\{ 
\left(\frac{\ds 2}{\ds 3}\sin^2\theta_W
\right)^2 \Bigl(Z\bar{u}(x) + N\bar{d}(x)\Bigr)
+ \left(\frac{\ds 1}{\ds 3}\sin^2\theta_W \right)^2 
\Bigl(Z\bar{d}(x) + N\bar{u}(x)\Bigr) \nonumber \\
\nms&+&\nms
\left(\frac{\ds 1}{\ds 2} - \frac{\ds 2}{\ds 3}
\sin^2\theta_W \right)^2 \Bigl(Z u(x) + Nd(x)\Bigr)
+ \left(- \frac{\ds 1}{\ds 2} + \frac{\ds 1}{\ds 3}\sin^2\theta_W \right)^2 
\Bigl(Z d(x) + Nu(x)\Bigr)
\Biggr\}  \Biggr] \ts,
\label{eq:DIntrl1} \\
\frac{\ds d^2 \sigma_{\rm nc}^{\bar{\nu}}}{\ds dx dy}
\nms&=& \Bigl(\mbox{exchange the factor $(1-y)^2$ with the factor 1 
in the above expression (\ref{eq:DIntrl1})} \Bigr) \ts, \quad\quad
\label{eq:DIntrl2}
\end{eqnarray}
where $G_{\rm F},\ts \theta_W,\ts m_{\rm N}$ and $E_{\nu}$ stand for the Fermi 
coupling constant, the Weinberg angle, nucleon mass and incident neutrino 
energy. $u(x)$ and $d(x)$ are the parton distribution densities for $u$ and 
$d$ quarks. $Z$ and $N$ stand for the proton number and the neutron number 
inside a given nuclei, and $q$ denotes the momentum transfer for the processes
considered. Here the variables $x$ and $y$ are defined as
\begin{eqnarray*}
x = \frac{\ds -q^2}{\ds 2m_{\rm N} \left( E_{\nu} - E_{\mu} \right)} \ts,\quad
y = \frac{\ds \left( E_{\nu} - E_{\mu} \right)}{\ds E_{\nu}} \es,
\end{eqnarray*}
where $E_{\mu}$ stands for the energy of  the muon in the final state.
In eqs.~(\ref{eq:DIchrgd1}) and (\ref{eq:DIchrgd2}) we take the following
scattering processes into consideration at the tree level:
\begin{eqnarray*}
\nu_{\mu} + d \to \mu^{-} + u \ts, \quad
\nu_{\mu} + \bar{u} \to \mu^{-} + \bar{d} \ts, \\
\bar{\nu}_{\mu} + u \to \mu^{+} + d \ts, \quad
\bar{\nu}_{\mu} + \bar{d} \to \mu^{+} + \bar{u} \ts, 
\end{eqnarray*}
Here we ignored the contribution of strange quark for simplicity.
Similarly, in eqs.~(\ref{eq:DIntrl1}) and (\ref{eq:DIntrl2}) we include the 
following neutral current scatterings:
\begin{eqnarray*}
\nu_{\mu} + {\rm q} \to \nu_{\mu} + {\rm q} \ts, \quad
\nu_{\mu} + \bar{{\rm q}} \to \nu_{\mu} + \bar{{\rm q}} \ts, \\
\bar{\nu}_{\mu} + {\rm q} \to \bar{\nu}_{\mu} + {\rm q} \ts, \quad
\bar{\nu}_{\mu} + \bar{{\rm q}} \to \bar{\nu}_{\mu} + \bar{{\rm q}} \ts,
\end{eqnarray*}
where ${\rm q} = u \mbox{ or } d$, and again we do not include the $s$ quark 
contribution at all.

  Second, we adopt a set of parton distributions called CTEQ5L~\cite{CTEQ} 
as the structure functions of proton and neutron. Using the set, we calculate 
the total cross sections $\sigma_{\nu p}^{\rm tot},\ts \sigma_{\nu n}^{\rm tot}
,\ts \sigma_{\bar{\nu} p}^{\rm tot}$ and $\sigma_{\bar{\nu} n}^{\rm tot}$
for the $\nu_{\mu} p,\ts \nu_{\mu} n,\ts \bar{\nu}_{\mu} p$ and 
$\bar{\nu}_{\mu} n$ scattering processes, respectively. Actually, we have 
carried out these calculations by 
the Monte-Carlo method with a high precision of $0.1~\%$.

  Finally, we obtain the mean free path of neutrino $R_{\nu}$ and that of
anti-neutrino $R_{\bar{\nu}}$ in a usual manner. For the purpose we first must 
know the number density of protons, $N_p$, inside the earth and that of 
neutrons, $N_n$. Indeed, we can easily obtain these densities, because we 
know that the number of protons and that of neutrons are on the average in the 
ratio of $49.5$ to $50.5$. In addition, the average mass density of the earth, 
$\rho_{\rm earth}$, is measured to be $5.52 \times 10^{3}~({\rm kg}/
{\rm m}^3)$. Hence, one can obtain the number densities:
\begin{eqnarray*}
N_p &=& \frac{\ds \rho_{\rm earth}}{\ds m_p} 
\times \frac{\ds 49.5}{\ds 49.5+50.5} \es (1/{\rm m}^3) \es, \\
N_n &=& \frac{\ds \rho_{\rm earth}}{\ds m_n} 
\times \frac{\ds 50.5}{\ds 49.5+50.5} \es (1/{\rm m}^3) \es,
\end{eqnarray*}
where $m_p$ and $m_n$ stand for the proton and the neutron masses.
Thus, we obtain the mean free path of (anti-) neutrino
\begin{eqnarray}
R_{\nu} &=& \frac{\ds 1}{\ds N_p\sigma_{\nu p}^{\rm tot} +
N_n \sigma_{\nu n}^{\rm tot}} \es ({\rm m}) \es, \\
R_{\bar{\nu}} &=& \frac{\ds 1}{\ds N_p\sigma_{\bar{\nu} p}^{\rm tot} +
N_n \sigma_{\bar{\nu} n}^{\rm tot}} \es ({\rm m}) \es.
\label{eq:mnfrpthNu}
\end{eqnarray}

  Throughout the process of the above calculations, we did not include higher
order corrections at all. However, it is plausible that if we include $s$ and
other heavier quarks into the parton distributions, the total cross sections
will probably become a few times larger than those obtained here. Furthermore, 
QCD corrections and the ``regeneration processes''~\cite{Naumov}
will also have other effects on the cross sections. Our next 
step is to take these effects into consideration on the basis of a detailed 
Monte-Carlo study, which remains to be solved in the near future.
\vspace*{0.1cm}
%
\section{Possible accelerator scheme}\label{sect:acclrtrNewTch}
\begin{figure}[h]
\vspace{0.8cm}
\begin{center}
\setlength{\unitlength}{1cm}
\begin{picture}(14.0,7.0)(0,0)
\put(0.5,0.0){\includegraphics[height=7cm]{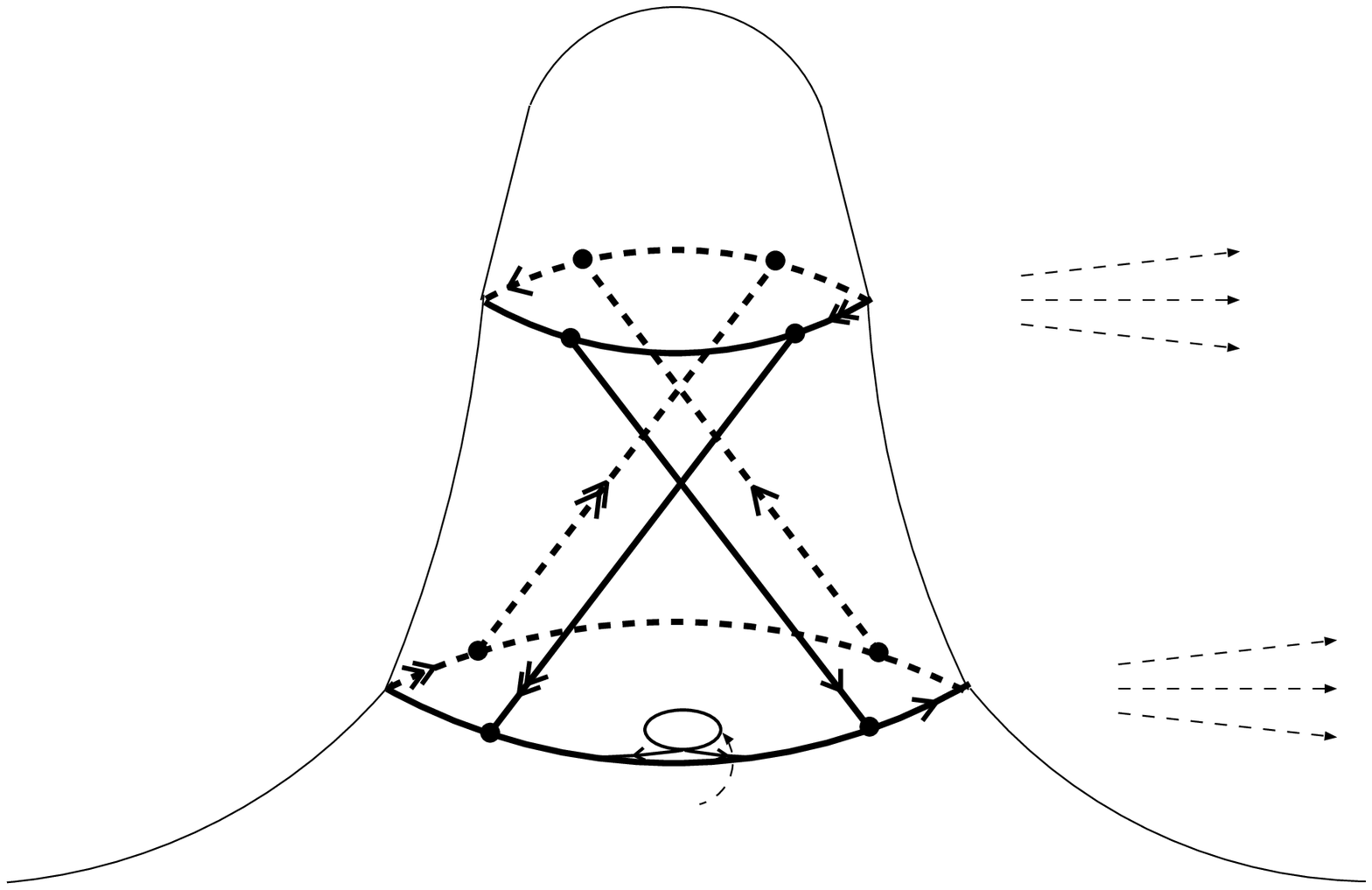}}
\put(10.6,4.6){hazardous plane} \put(11.4,1.4){hazardous plane}
\put(0.7,1.4){synchrotron $A$} \put(1.4,4.65){synchrotron $B$}
\put(4.8,0.2){injection system}
\put(4.7,5.1){\footnotesize $P_4$} \put(6.7,5.1){\footnotesize $Q_4$}
\put(4.7,4.0){\footnotesize $P_1$} \put(6.8,4.05){\footnotesize $Q_1$}
\put(3.90,2.05){\footnotesize $Q_3$} \put(7.5,1.95){\footnotesize $P_3$}
\put(4.0,0.89){\footnotesize $Q_2$} \put(7.28,0.9){\footnotesize $P_2$}
\end{picture}
\end{center}
\vspace*{-0.5cm}
\caption
{{\sl 
Accelerator scheme.
}}
\label{fig:acclrtrNewTch}
\vspace{0.2cm}
\end{figure}

\spc We first look for a mountain like in fig.~\ref{fig:acclrtrNewTch} whose 
surface does not touch many of the straight lines depicted as $P_1P_2,\ts
P_3P_4, \ts Q_1Q_2$ or $Q_3Q_4$. We construct two synchrotron $A$ and $B$
which are both revolvable. $A$ should be larger than $B$. Muon beam is injected
into the synchrotron $A$ first and accelerated to a sufficient energy.
Injection system could be installed in a tunnel in the mountain. Then it is 
stored either in the path $P_2P_3P_4P_1P_2$ or $Q_2Q_3Q_4Q_1Q_2$ depending on
the direction of the beam in the synchrotron $A$. The beam is either $\mu^{+}$
or $\mu^{-}$. The straight sections $P_1P_2,\ts P_3P_4, \ts Q_1Q_2$ and 
$Q_3Q_4$ are made of chambers separated by many bellow structures so that they
can have a flexible length. We probably have to prepare several chambers to 
cover from the minimum to the maximum length continuously. When we rotate
$A$ or $B$ the chambers must follow until we steer the straight section to
a given target.The next question is how precisely we can steer it. From the
discussion given in the text the required accuracy is $10^{-7}$. This is
$1/10$ micron per meter. We believe this is not an outrageous number. The
current effort toward the construction of a linear collider is aiming at 
approximately 1~micron per meter.
Future technology certainly will reach our required number sooner or later.%

  Another issue is the power consumption and the radiation hazard. Power 
required is $10^{14} \times 10^{-19} \times 10^{15}~{\rm W} \simeq 
10~{\rm GW}$. Actually, we may need something like 50~GW (considering the
efficiency) which is exactly the whole capacity of Japanese nuclear power.
But the energy consumption could be as small as $10^{2}/10^{8} = 10^{-6}$
times the whole consumption of 50~GW power. This should be quite tolerable.
For the radiation hazard we have two planes in fig.~\ref{fig:acclrtrNewTch}
which should not be crossed by anyone during the operation and one direction
toward the sky where no one is allowed to touch. The other direction is,
of course, toward the target. Almost all the energy is lost in the earth
and only $10^{-7}$ times the whole energy hits the target. People working
near the target should be warned unless they are working to conceal the
weapons.

  We can perform $\mu^{+}\mu^{+},\ts \mu^{-}\mu^{-}$ and $\mu^{+}\mu^{-}$
colliding experiment in this scheme by injecting two beams simultaneously
although the detector should be placed on a very steep slope between $A$ and
$B$ synchrotrons. We believe it is not unreasonable to build this kind of
accelerator complex first with much lower energy beam to study the inside of 
the earth and simultaneously performing the muon collider experiments and also
the neutrino experiments.
\vspace*{1.0cm}

\noindent{\Large \bf Acknowledgment}

  We are grateful to Tetsuo Abe for giving us fig.~\ref{fig:mfpVsEng} of the
mean free paths. We thank Masayoshi Kawai for his useful comments on MCNPX. 
We appreciate Yoshinobu Takaiwa for his helpful suggestion of some Monte-Carlo 
generators. We would like to acknowledge valuable discussions with 
Prof.~Tokushi Shibata and members of Theory Division of KEK. The questions and 
comments by Yoshitaka Kimura, Sakuei Yamada, Frank~von Hippel, Sydney Drell, 
Burton Richter and colleagues of H.S.~at the University of Hawaii are 
deeply appreciated.
\vspace*{0.5cm}
%
%


\begin{thebibliography}{99}
\bibitem{FTPR} 
Our calculation is now in progress on the basis of the algorithm adopted 
in the generator HERWIG or MARS. In addition, we must take the neutrino
transport theory into consideration. The result will be reported in the near 
future.
\end{thebibliography}
\end{document}